\documentclass[a4paper,
              ]{jacow}
\makeatletter%
	\ifboolexpr{bool{xetex}}
	 {\renewcommand{\Gin@extensions}{.pdf,%
	                    .png,.jpg,.bmp,.pict,.tif,.psd,.mac,.sga,.tga,.gif,%
	                    .eps,.ps,%
	                    }}{}
\makeatother

\ifboolexpr{bool{xetex} or bool{luatex}} % test for XeTeX/LuaTeX
 {}                                      % input encoding is utf8 by default
 {\usepackage[utf8]{inputenc}}           % switch to utf8

\usepackage[USenglish]{babel}			

\usepackage[final]{pdfpages}
\usepackage{multirow}
\usepackage{ragged2e}

\ifboolexpr{bool{jacowbiblatex}}%
 {%
  \addbibresource{jacow-test.bib}
  \addbibresource{biblatex-examples.bib}
 }{}
\begin{document}

\title{Design, fabrication and high-gradient tests of X-band choke-mode structures
\thanks{Work supported by the National Natural Science Foundation of China (Grant No. 11135004).}}

\author{Xiaowei Wu\thanks{wuxw12@mails.tsinghua.edu.cn}\textsuperscript{1}, Jiaru Shi\textsuperscript{1}, Huaibi Chen\textsuperscript{1}, Hao Zha\textsuperscript{1}
\\ Department of Engineering Physics Tsinghua University, Beijing CN-100086, China\\
		\textsuperscript{1}also at Key Laboratory of Particle Radiation Imaging, Ministry of Education, Beijing, China\\
		\\Tetsuo Abe, Toshiyasu Higo, Shuji Matsumoto
\\KEK, High Energy Accelerator Research Organization, Tsukuba, 305-0801, Japan}	

\maketitle

\begin{abstract}
   Two standing-wave single-cell choke-mode damped structures with different choke dimensions which worked at 11.424 GHz were designed, manufactured and tuned by accelerator group in Tsinghua University. High power test was carried out to study choke-mode structure's properties in
   high gradient and related breakdown phenomenon. A single-cell structure without choke which almost has the same inner dimension as choke-mode structure was also tested as a comparison to study how the choke affects high-gradient properties. In this paper, we report on the latest status of the high power test, including various observations and the experimental results.
\end{abstract}

\section{INTRODUCTION}
 As an alternative
design for CLIC main accelerating structures, X-band
choke-mode structures had been studied under the collaboration
between Tsinghua University, CERN and KEK~\cite{CLIC1,choke1,choke2,choke3}.
Three X-band single-cell standing-wave structures including
two choke-mode structures and one reference structure
without choke were designed, fabricated, assembled, and tuned by Tsinghua University. The high power
test, aiming at studying the high-gradient properties of
X-band choke-mode structure, were carried out at Nextef~\cite{nextef1} facility in KEK. One of the choke-mode structures was cut into three pieces for inner surface observation after the high power test. Below we report the main results of the test. Observations from post-mortem are also presented.

\section{overview OF THE SINGLE-CELL STRUCTURES}

%\subsection{RF design}

\begin{figure}[!htb]
%   \vspace*{-.5\baselineskip}
   \centering
   \includegraphics*[width=180pt]{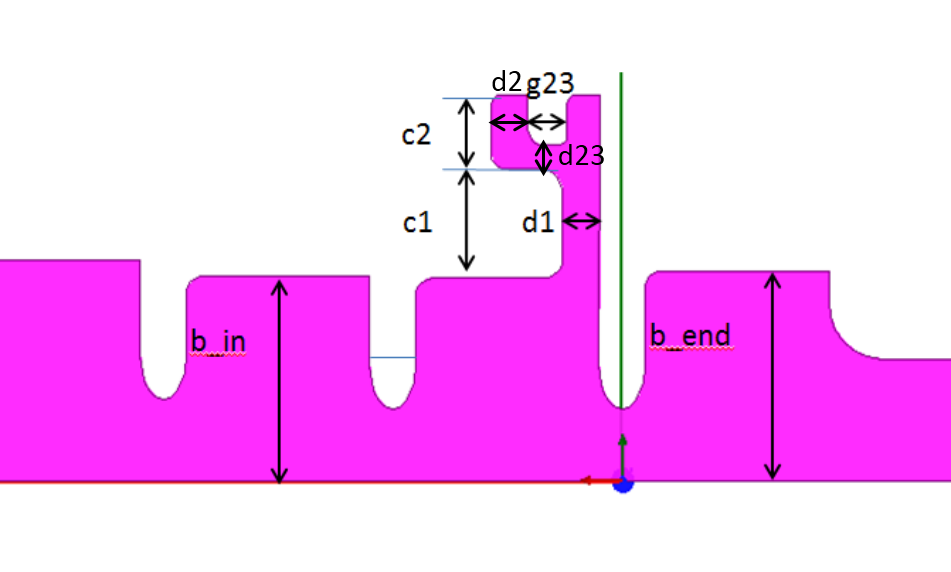}
   \caption{Choke-mode structure geometry.}
   \label{2-f2}
%   \vspace*{-\baselineskip}
\end{figure}

The single-cell standing-wave structure consists of three
parts: the input coupler cell, the high-gradient middle
cell(s), and the end cell~\cite{sc1}.
Three single-cell standing-wave structures including two
choke-mode designs and one reference design were proposed
by Tsinghua University. Geometry of the choke-mode structure is shown in Fig.~\ref{2-f2}.
The names of the single-cell structures are derived from the manufacturer’s name plus structure's type and key geometry.  An example of a single-cell structure name is: THU-CHK-D1.26-G1.68. Here THU is the manufacturer and CHK is the structure's type. D1.26 is the d23 dimension in mm and G1.68 is the d1 dimension in mm, as shown in Fig.~\ref{2-f2}. The single-cell structures we designed and tested are: THU-CHK-D1.26-G1.68, THU-CHK-D1.26-G2.1 and THU-REF\footnote[1]{Note the nomenclature here is different from that in~\cite{choke3}. Dimension of d23 shown in Fig.~\ref{2-f2} is added in the structure's name.}. The details of RF design can be found in \cite{choke3}.

\begin{figure}[!htb]
%   \vspace*{-.5\baselineskip}
   \centering
   \includegraphics*[width=180pt]{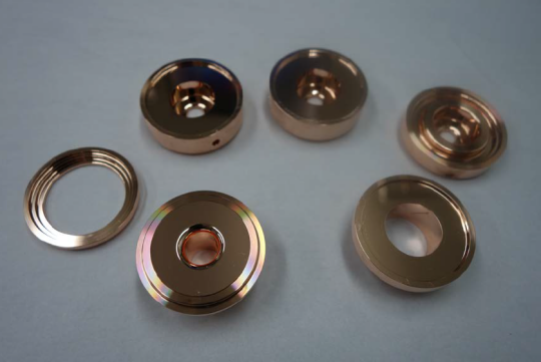}
   \caption{Disks of single-cell choke-mode strutures before bonding.}
   \label{2-f1}
%   \vspace*{-\baselineskip}
\end{figure}

The mechanical design of single-cell choke-mode structure was made
of 6 disks, as shown in Fig.~\ref{2-f1}. The middle cell with choke
was achieved by two disks together. All of these disks were
manufactured by turning because of the symmetrical design
of choke-mode structure.

 The disks were first cleaned and etched by the internal procedures based on GLC fabricating technology and then bonded. The contact areas of each choke-mode structure disks are not consistent vertically. This may cause deformations during the bonding. A choke-mode structure prototype was fabricated for diffusion bonding test to check the bonding quality. During the bonding test, the prototype was cut into halves for bonding contact area check. After completion of the successful bonding test, the individual parts of the single-cell structures were diffusion bonded in a hydrogen furnace at Tsinghua University. Operating
frequency was tuned to 11.424 GHz at the working temperature of
30 $^{\circ}$C which is the standard cooling water temperature at
Nextef. The structures were vacuum baked at 500 $^{\circ}$C for 5 days. The structures were kept under vacuum after baking by sealing with a valve and then shipped to KEK also under vacuum. Rf measurement results at KEK kept consistent with that tested in Tsinghua University.

\iffalse
If a displayed equation needs a number (i.\,e., it will be
referenced), place it it in parentheses, and flush with the
right margin of the column. The equation itself should be
indented and centred, as far as is possible:
\begin{equation}\label{eq:units}
    C_B=\frac{q^3}{3\epsilon_{0} mc}=\SI{3.54}{\micro eV/T}
\end{equation}

When referencing a numbered equation, use the word
“Equation” at the start of a sentence, and the abbreviated
form, “Eq.”, if in the text. The equation number is placed
in parentheses [e.g., Eq. (1)].
\fi

\section{High power test}
High power test was carried out after the structure was installed in Shield-B~\cite{nextef1} of Nextef at KEK. THU-CHK-D1.26-G1.68 was first tested followed by THU-CHK-D1.26-G2.1 and THU-REF.
Nextef, which stands for New
X-band Test Facility of KEK, was founded in 2006 as a
reassembled facility of GLCTA~\cite{nextef2,nextef3} as a 100 MW high
power station for X-band accelerating structure study. Shield-B is aiming at basic high-gradient study by testing single-cell structures~\cite{nextef4}.

\subsection{Test stand}
\begin{figure}[!htb]
%   \vspace*{-.5\baselineskip}
   \centering
   \includegraphics*[width=230pt]{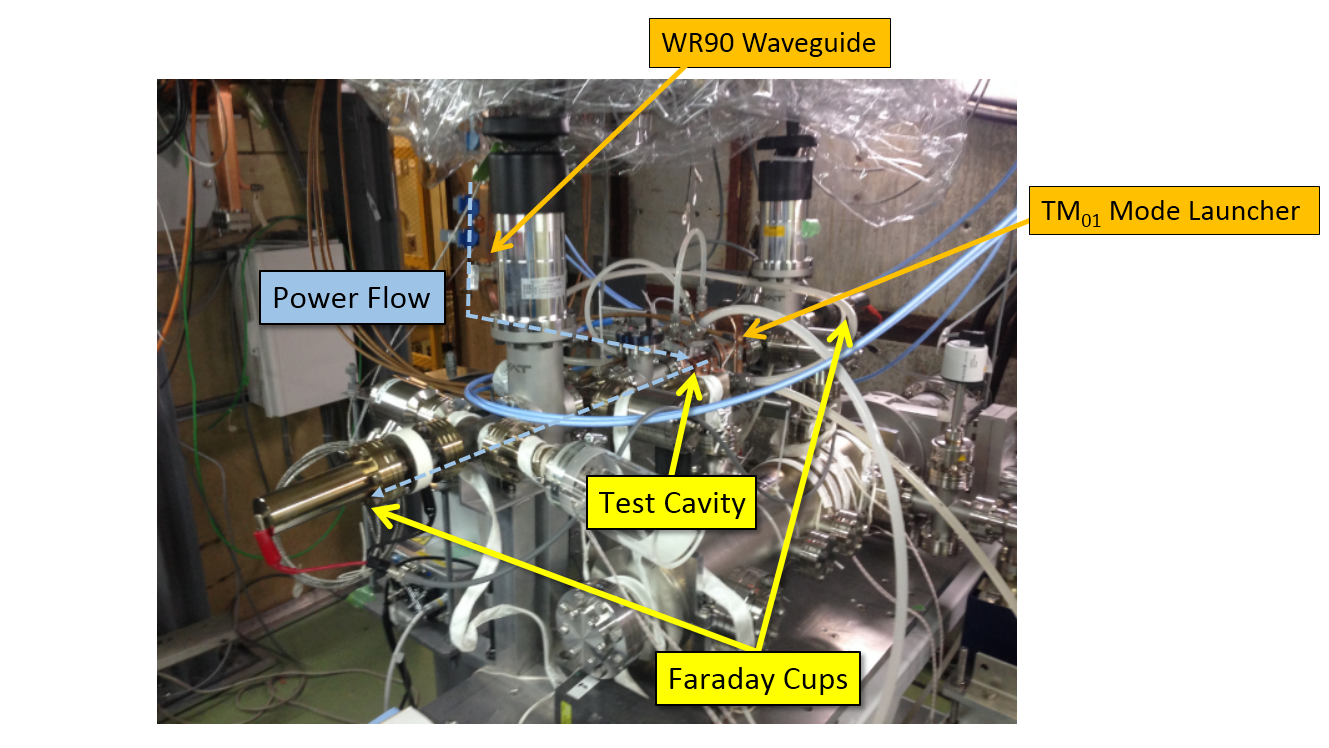}
   \caption{Experimental setup for the high power test of choke-mode structure.}
   \label{3-f1}
%   \vspace*{-\baselineskip}
\end{figure}

The experimental setup for high-gradient tests is shown in Fig.~\ref{3-f1}. RF power was transferred to Shield-B via WR90 waveguide and circular low-loss wave guide and then fed into the cavity from
the TM$_{01}$ mode launcher seen in the right side. The reflection rf
signal and the dark current signals were monitored pulse-by-pulse during the operation for breakdown
detection. Once breakdown occurred, we stopped the next rf pulse and waited for several tens of
seconds before the next rf pulse. Typically, the operation would reduce the rf power by about 5\% and ramp the power again by increasing 0.2 MW in 20 seconds.

\subsection{RF waveforms}
\begin{figure}[!htb]
%   \vspace*{-.5\baselineskip}
   \centering
   \includegraphics*[width=180pt]{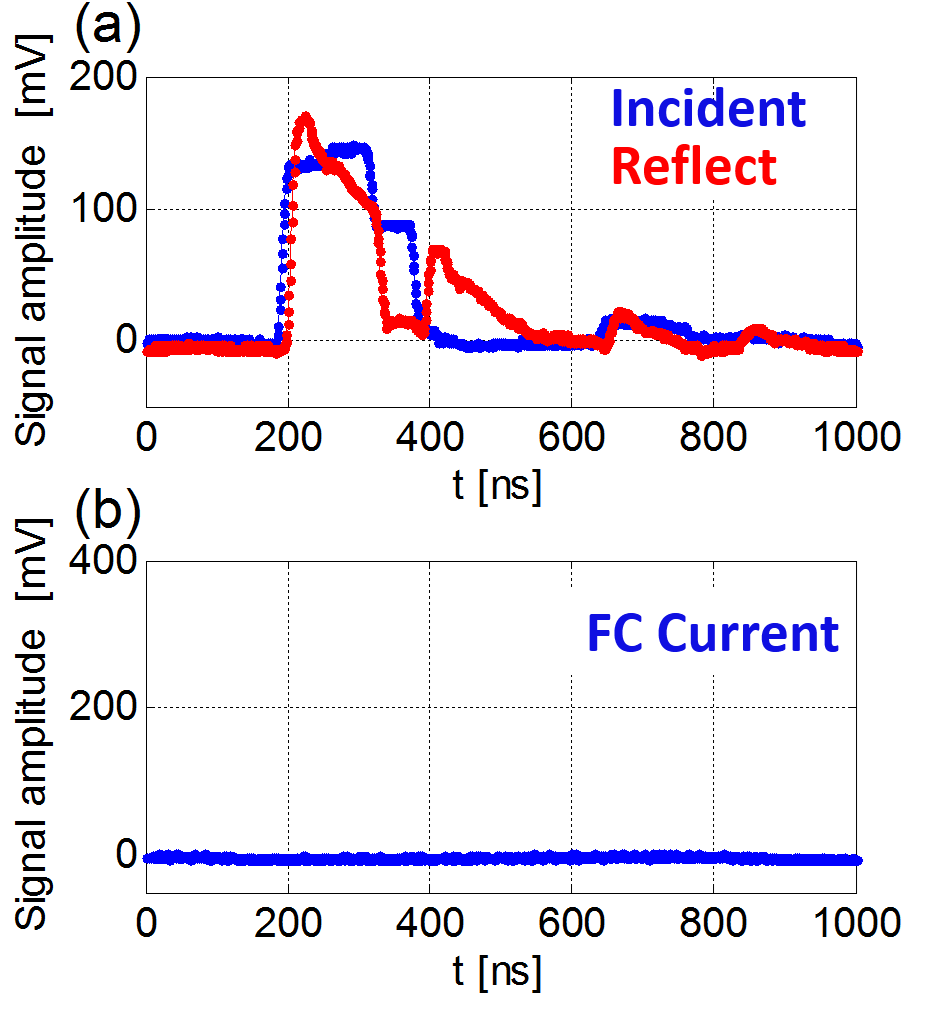}
   \caption{Normal event signals. (a) Incident wave and reflection wave. (b) Faraday cup signal.}
   \label{3-f3}
%   \vspace*{-\baselineskip}
\end{figure}

In order to have a constant high field in the test cell as
long as possible, we shaped the rf pulse into double steps as
shown in Fig.~\ref{3-f3}(a). In the first step for 100 ns, we charged
RF energy in the test cavity (charging step).
In the second step for 100 to 300 ns (maintaining step), the rf power was decreased to approximately 36\% of that in the charging step~\cite{nextef4}. The width of maintaining step was changed in different pulse width operations. The current signal collected by the Faraday cup is shown in Fig.~\ref{3-f3}(b).
\subsection{Experimental results}
\begin{figure}
%   \vspace*{-.5\baselineskip}
   \centering
   \includegraphics*[width=180pt]{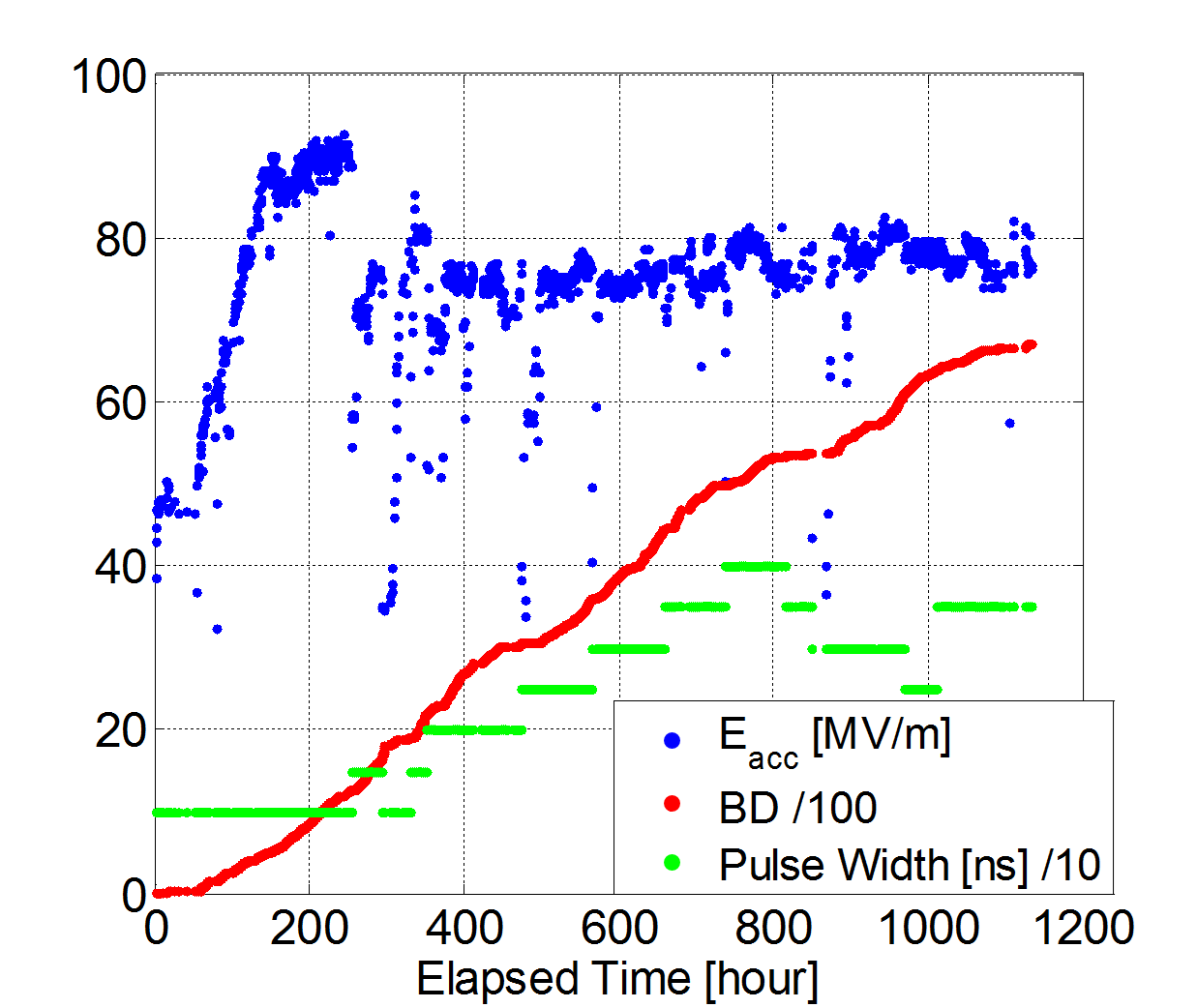}
   \caption{High-gradient testing history of THU-CHK-D1.26-G1.68. The blue
dots are the E$_{acc}$ [MV/m], the green dots are the pulse
width [ns] divided by ten and the red dots are breakdown
number divided by 100.}
   \label{3-f2}
%   \vspace*{-\baselineskip}
\end{figure}

The summary of the conditioning history of THU-CHK-D1.26-G1.68 is shown in
Fig.~\ref{3-f2}. The blue, green and red points represent the Eacc,
the pulse width of rf power and the accumulated number
of breakdowns, as a function of elapsed hours
respectively. The E$_{acc}$ value was recorded at every
interlock event. The dots which fall below the envelope of E$_{acc}$
correspond to interlocks in the power ramping stage after
previous breakdown. Rf power could not be further increased after 100 hours in 100 ns pulse width operation due to continuous breakdowns. Same phenomenon happened at longer pulse width operation. The maximum gradient obtained in the test was 75 MV/m as shown in Fig.~\ref{3-f2}.

\begin{figure}
    \centering
    \includegraphics*[width=180pt]{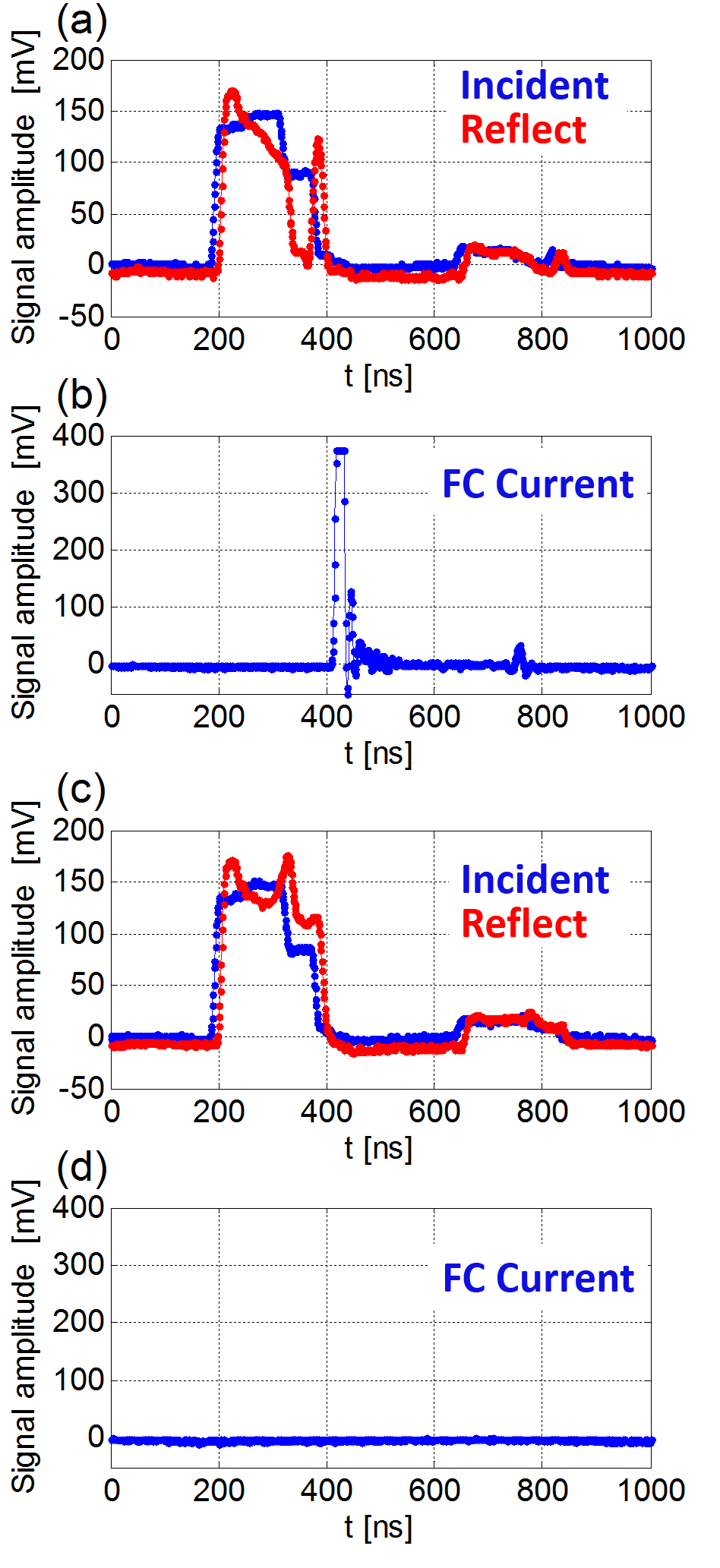}
    \caption{Examples of two types of breakdowns. (a) and (b) are incident and reflection waves and Faraday cup signal of the breakdown accompanied with current flash. (c) and (d) are the signals of the breakdown accompanied without current flash.}
    \label{3-f4}
\end{figure}

Two types of breakdowns, which were accompanied with and without current flash, were observed in the high power test. Waveforms of these two breakdowns are shown in Fig.~\ref{3-f4}. Breakdowns were accompanied with the current flash into the Faraday cup during the initial ramping stage. After initial ramping, few current flash breakdowns were observed in the detected events. As the Faraday cups were located at the end of the pipe axis, the electrons emitted from the choke breakdown area were not easily collected. It was speculated that breakdowns with current flash happened in the cylinder cavity while breakdowns without current flash happened in the choke. Frequent breakdowns in the choke during the high-gradient test were assumed to be the main limitation of obtaining higher gradient as shown in Fig.~\ref{3-f2}. This speculation was verified in the post-mortem observation. This will be discussed in the next section.

\section{Post-mortem observation}

\begin{figure}[!htb]
%   \vspace*{-.5\baselineskip}
   \centering
   \includegraphics*[width=190pt]{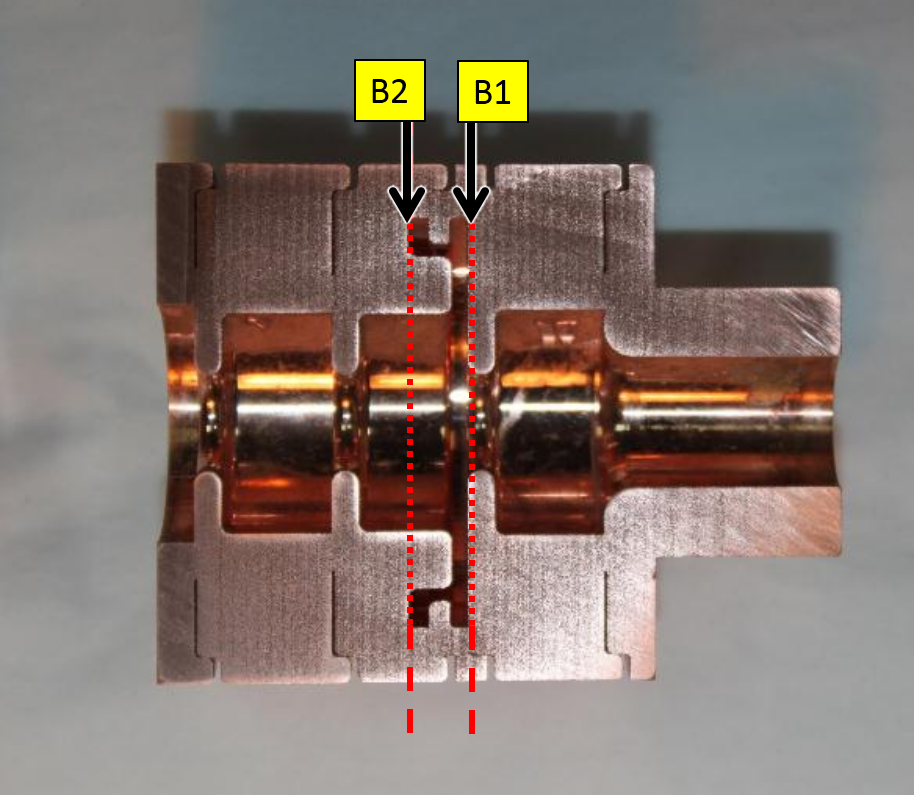}
   \caption{Cutting plan of THU-CHK-D1.26-G1.68 (cross section view). This is the photo of the choke-mode structure prototype made for bonding test which was cut into halves along axial direction. THU-CHK-D1.26-G1.68 was cut twice along B1 and B2 lines. }
   \label{4-f1}
%   \vspace*{-\baselineskip}
\end{figure}

THU-CHK-D1.26-G1.68 was cut after the high-gradient test was finished. The structure was cut along radial direction twice as shown in Fig.~\ref{4-f1}. Cutting along radial direction allowed microscopy imaging of the choke surfuce which was speculated as frequent breakdown sites. The surfaces of the irises and cylinder cavity are very clean while the surfaces of the choke groove are very rough with naked eye observation. The structure was then examined with a microscope. The microscopy's model is KEYENCE VE-8800.

\begin{figure}[!htb]
%   \vspace*{-.5\baselineskip}
   \centering
   \includegraphics*[width=190pt]{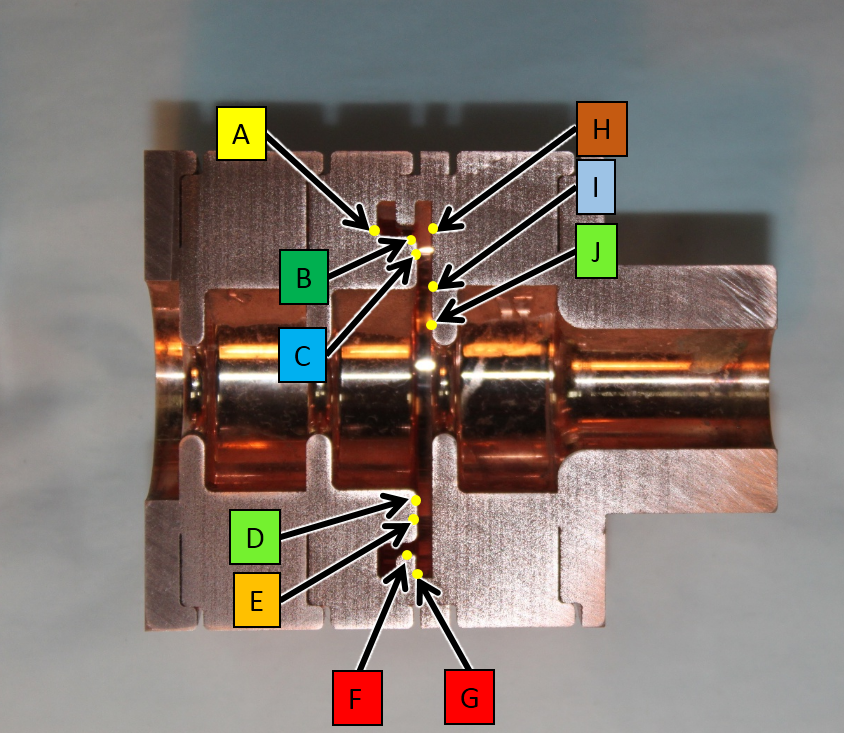}
   \caption{Inner surface inspecting points of THU-CHK-D1.26-G1.68. }
   \label{4-f2}
%   \vspace*{-\baselineskip}
\end{figure}

Ten points were chosen for inner surface inspections as show in Fig.~\ref{4-f2}.

\begin{figure*}[t]
%   \vspace*{-.5\baselineskip}
   \centering
   \includegraphics*[width=\textwidth]{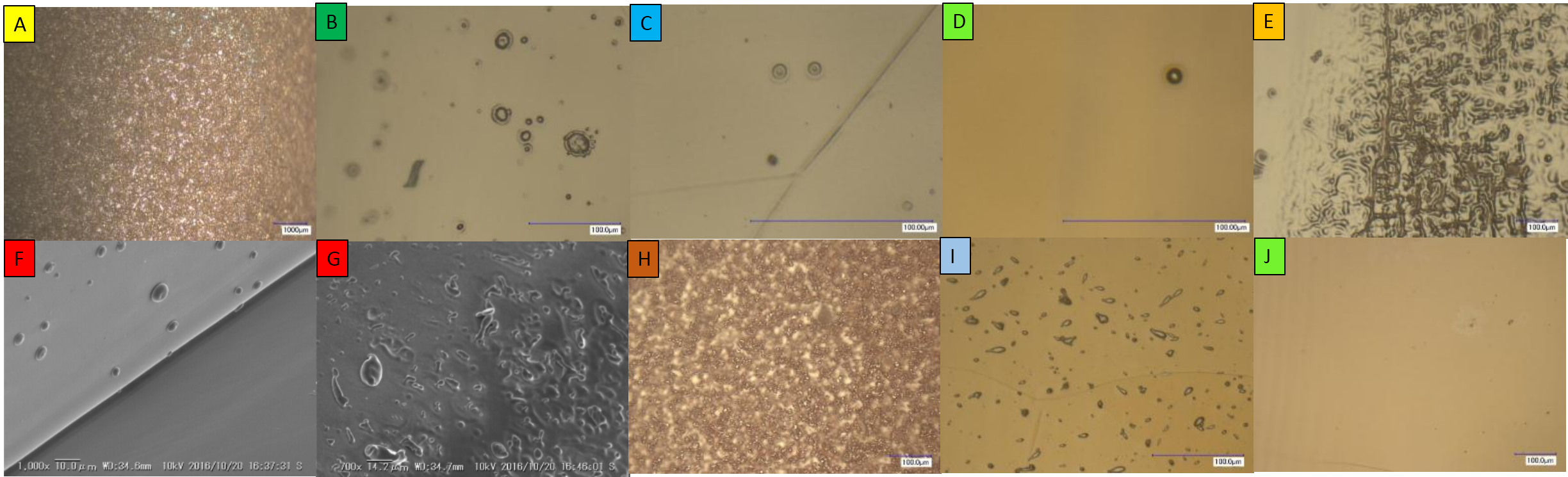}
   \caption{Results of the inner surface inspections. }
   \label{4-f3}
%   \vspace*{-\baselineskip}
\end{figure*}

The inner surface observation results are shown in Fig.~\ref{4-f3}.  The areas of cylinder cavity and irises are clean as shown in point D and J. This indicates that few breakdowns occurred in these areas. Microscopy imaging of piont B, E, F, G, H and I showed damage such as "craters", small "protrusions" and "speckles". Point B and F showed significantly more damage than the other points. The damage at point E, G, H and I is speculated as melting copper sputtered from the choke area. Therefore, d2 area shown in Fig.~\ref{2-f2} has a high breakdown rate.
\iffalse
\begin{figure*}[!b]
%   \vspace*{-.5\baselineskip}
   \centering
   \includegraphics*[width=\textwidth]{choke_field}
   \caption{Field information of THU-CHK-D1.26-G1.68. }
   \label{4-f4}
%   \vspace*{-\baselineskip}
\end{figure*}
\fi

\section{status and plan}
The high-gradient tests of three structures have been finished in 2017.1. However, the data of the rest two structures are still under analyzing. In order to achieve higher gradient in choke-mode structure, we have designed new single-cell choke-mode structures with larger d2. New structures are under fabrication for the high-gradient test.

\section{Conclusion}
   Two standing-wave single-cell choke-mode damped structures as well as one reference structure have been successfully designed and fabricated at Tsinghua University.
High power tests were carried out at the Nextef facility in KEK and the test
demonstrated that the present choke-mode structure can operate at a highest gradient of 75 MV/m.
Two types of breakdowns, which were accompanied with and without current flash, were observed in the test. The former one was speculated to be the breakdown occurred in the iris and cylinder cavity area while the latter one was speculated to be located in the choke. Post-mortem of THU-CHK-D1.26-G1.68 verified this speculation and indicated that d2 area of the choke is the critical limitation of obtaining higher gradient. The present d2 with dimension of 1.26 mm will cause continuous breakdowns around 75 MV/m.

\section{acknowledgement}
This work was supported by the National Natural
Science Foundation of China (Grant No. 11135004). The
experimental program had also been supported as one of
the collaborations of the CLIC under the agreement
between Tsinghua University and CLIC and that,
ICA-JP-0103, between KEK and CERN. The authors
thank those of KEK electron-positron injector group for
supporting the long-term operation.

\iffalse  % only for "biblatex"
	\newpage
	\printbibliography

% "biblatex" is not used, go the "manual" way
\else

%\begin{thebibliography}{99}   % Use for  10-99  references

\null  % this is a hack for correcting the wrong un-indent by package 'flushend' in versions before 2015
\end{document}